\documentclass[apj]{emulateapj}

\usepackage{graphicx}

\newcommand {\Lya}  {Ly$\alpha$}  

\newcommand {\HI}   {\ion{H}{1}}  
\newcommand {\HeII} {\ion{He}{2}}  

\newcommand {\OI}   {\ion{O}{1}}  
\newcommand {\OVI}  {\ion{O}{6}}  

\newcommand {\CII}  {\ion{C}{2}}

\newcommand {\CIV}  {\ion{C}{4}}

\newcommand {\NI}   {\ion{N}{1}}
\newcommand {\NV}   {\ion{N}{5}}

\newcommand {\SiIV}  {\ion{Si}{4}}
\newcommand {\SiIII} {\ion{Si}{3}}
\newcommand {\SiII}  {\ion{Si}{2}}

\newcommand {\AlII}  {\ion{Al}{2}}
\newcommand {\SII}  {\ion{S}{2}}
\newcommand {\PII}  {\ion{P}{2}}

\newcommand {\FeII}  {\ion{Fe}{2}}
\newcommand {\kms}  {km~s$^{-1}$}

\newcommand {\HST} {{\it HST}}

\newcommand {\etal} {et~al.}

\newcommand{\fluxunits}{erg cm$^{-2}$ s$^{-1}$ \AA$^{-1}$}

\slugcomment{\sc{Accepted to ApJ :} September 23, 2011}
\shorttitle{The Cosmic Origins Spectrograph}
\shortauthors{Green \etal}

\begin{document}
\title{The Cosmic Origins Spectrograph}
\author{James C. Green\altaffilmark{1,2}, Cynthia S. Froning\altaffilmark{2}, Steve Osterman\altaffilmark{2},
Dennis Ebbets\altaffilmark{3}, Sara H. Heap\altaffilmark{4}, Claus Leitherer\altaffilmark{5} Jeffrey L. Linsky\altaffilmark{6},
Blair D. Savage\altaffilmark{7}, Kenneth Sembach\altaffilmark{5}, J. Michael Shull\altaffilmark{1,2}, Oswald H.W. Siegmund\altaffilmark{8},
Theodore P. Snow\altaffilmark{1,2}, John Spencer\altaffilmark{9}, S. Alan Stern\altaffilmark{9},
John Stocke\altaffilmark{1,2}, Barry Welsh\altaffilmark{10},
St\'{e}phane B\'{e}land\altaffilmark{2}, Eric B. Burgh\altaffilmark{2}, Charles Danforth\altaffilmark{2}, Kevin France\altaffilmark{2}, Brian Keeney\altaffilmark{2}, Jason McPhate\altaffilmark{10}, Steven V. Penton\altaffilmark{2},
John Andrews\altaffilmark{9}, Kenneth Brownsberger\altaffilmark{3}, Jon Morse\altaffilmark{11}, and Erik Wilkinson\altaffilmark{9}}

\altaffiltext{1}{Department of Astrophysical and Planetary Sciences, University of Colorado, 391-UCB, Boulder, CO 80309}
\altaffiltext{2}{Center for Astrophysics and Space Astronomy, University of Colorado, 389-UCB, Boulder, CO 80309}
\altaffiltext{3}{Ball Aerospace \& Technologies Corp., 1600 Commerce Street, Boulder, CO 80301}
\altaffiltext{4}{NASA Goddard Space Flight Center, Code 681, Greenbelt, MD 20771}
\altaffiltext{5}{Space Telescope Science Institute, 3700 San Martin Drive, Baltimore, MD 21218, USA}
\altaffiltext{6}{JILA, University of Colorado and NIST, Boulder, CO 80309-0440, USA}
\altaffiltext{7}{University of Wisconsin-Madison, 475 North Charter Street, Madison, WI 53706, USA}
\altaffiltext{8}{Astronomy Department, University of California, Berkeley, CA 94720 USA}
\altaffiltext{9}{Southwest Research Institute, 1050 Walnut St, Suite 300, Boulder, CO 80302}
\altaffiltext{10}{Space Sciences Laboratory, University of California, 7 Gauss Way, Berkeley, CA, 94720, USA}
\altaffiltext{11}{NASA Headquarters, Washington, DC 20546-0001, USA}

\begin{abstract}
The Cosmic Origins Spectrograph (COS) is a moderate-resolution spectrograph with unprecedented sensitivity that was installed into the \emph{Hubble Space Telescope} (\HST) in May 2009, during \HST\ Servicing Mission 4 (STS-125).  We present the design philosophy and summarize the key characteristics of the instrument that will be of interest to potential observers. For faint targets, with flux $F_{\lambda} \approx 1.0 \times10^{-14}$ \fluxunits, COS can achieve comparable signal to noise (when compared to STIS echelle modes) in 1-2\% of the observing time.  This has led to a significant increase in the total data volume and data quality available to the community. For example, in the first 20 months of science operation (September 2009 - June 2011) the cumulative redshift pathlength of extragalactic sight lines sampled by COS is 9 times that sampled at moderate resolution in 19 previous years of \emph{Hubble} observations. COS programs have observed 214 distinct lines of sight suitable for study of the intergalactic medium as of June 2011.  COS has measured, for the first time with high reliability, broad \Lya\ absorbers and Ne VIII in the intergalactic medium, and observed the \HeII\ reionization epoch along multiple sightlines. COS has detected the first CO emission and absorption in the UV spectra of low-mass circumstellar disks at the epoch of giant planet formation, and detected multiple ionization states of metals in extra-solar planetary atmospheres. In the coming years, COS will continue its census of intergalactic gas, probe galactic and cosmic structure, and explore physics in our solar system and Galaxy.
\end{abstract}

\keywords{spectrographs, ultraviolet:general}

\section{Introduction}\label{sec:intro}

The Cosmic Origins Spectrograph (hereafter COS) was proposed as an instrument for the \emph{Hubble Space Telescope}
(\HST) Servicing Mission 4 (SM4) in response to a NASA Announcement of Opportunity released in 1996 (AO-96-OSS-03).
It was conceived as a point-source, moderate-resolution spectrograph, with significantly greater observing sensitivity than
previous or planned instruments and a select number of operating modes. The primary science case was the
extension of the number of QSO absorption sight lines from the handful available, even after the installation
of STIS on SM2, to several hundred, to begin the systematic study of the distribution and physical conditions
of the intergalactic medium (IGM) and the structure of matter on cosmological scales.

In fall 1997, COS was the only proposal selected for development, with a presumed launch date of October 2002.
At the time of selection, \HST\ was scheduled for decommissioning in 2005, so that the proposed version of COS
was intended for only three years of operation, consistent with the guidelines of the announcement of opportunity.
The original proposal for COS included only the short wavelength, Far Ultraviolet (FUV) channels, and covered
1150--1775 \AA. After selection, NASA directed the project to incorporate a Near Ultraviolet (NUV) capability and extend the
wavelength coverage out to 3200 \AA. The channels that support the 1750--3200 \AA\ regime are referred to as
the NUV channels. In 1998, the planned lifetime of \HST\ was extended to 2010, and the Wide
Field Camera 3 (WFC3) was added as a second instrument for SM4. The extension of the mission lifetime required a significant
change in the electronics parts program of COS to provide additional radiation protection, and the upgrade of numerous
components to full redundancy, which had previously been viewed as unnecessary for a three-year mission lifetime.
In November 2003, COS completed thermal vacuum testing at Ball Aerospace and was being prepared for delivery for flight. However,
 SM4 was cancelled by NASA in 2004. In the aftermath of the Space Shuttle Columbia disaster, it was considered too
risky to fly a mission that could not utilize the International Space Station as a safe haven in case of damage to the Shuttle.

SM4 was reinstated in 2006, with a scheduled launch date of 2008. After a slip to accommodate
the servicing of one of the \HST\ command and data handling computers (which failed in 2008), launch and servicing occurred
in May 2009. After orbital verification and initial calibration, COS began science operations in September 2009.

The COS Instrument Handbook \citep{Dixon10}, available online at the Space Telescope Science Institute (STScI) web site\footnote{http://www.stsci.edu/hst/cos/documents/handbooks/current}, provides constantly
updated information about the on-orbit performance of COS.  While some overlap exists between this work and those
documents, it is our intention to primarily discuss those aspects of the COS instrument that are not included in
the STScI documentation. Potential observers are advised to refer to the current version of the Handbook for
the latest information on COS before submitting a proposal.

\section{Design Philosophy}\label{sec:design}

COS was envisioned as a high-throughput, moderate-resolution spectrograph whose high level of sensitivity would
enable the efficient observations of large numbers of targets, united by the theme of probing cosmic origins.
The centerpiece of this vision was a sufficiently large survey of absorption lines, using QSOs as background sources,
so as to greatly advance
our understanding of the structure of matter on cosmological scales, in particular, the physical conditions,
metallicity, and distribution of the cosmic web. We established a design goal of being ten times more efficient
(1/10$^{th}$ the observing time) at obtaining the same S/N spectra for a given object, compared to the predicted
performance of STIS. At the time the COS concept was being developed, Servicing Mission 2 had not yet occurred (SM2 was conducted in February 1997) and
STIS was not yet installed. The final performance of COS substantially exceeds this design goal, as will be
discussed in Section 6. A factor-of-ten improvement in observing efficiency would make almost
1000 times as many QSO lines of sight available for observation in a reasonable time, many more than could
possibly observed by \HST\ during its final years. Thus, the QSO survey would be limited only by the amount
of observing time that would be assigned to this science objective, and not by a lack of observable targets.
In addition, this large increase in potential targets means that observing programs could have greater
freedom to choose their lines of sight, and could design probes of particularly interesting foreground regions,
rather than being limited to the serendipitous sampling of the foreground provided by the handful of distant background
targets that
were bright enough to observe prior to COS.

To this end, the spectral resolving power of COS ($R ~ = ~ \lambda / \Delta \lambda$) was chosen to be 20,000 in the
primary observing mode, as this was felt to be sufficient to avoid confusion between intergalactic absorbers,
and to allow reasonable estimation of the physical parameters of the gas. \Lya\ and \OVI\ absorption lines have typical doppler
$b$ values of 20 - 25 km s$^{-1}$ and the observed \Lya\ line frequency $dN/dz \sim 100$ down to an equivalent width of 20 m\AA, yielding one line about every 300  km s$^{-1}$.
While higher spectral resolution could
have been employed, this would have required more observing time to achieve the same signal to noise per
resolution element and restricted the simultaneous bandpass coverage.  Given the resolving power constraint,
all other optical design parameters for COS were chosen to minimize the time necessary to achieve a given signal to noise.
It was our intention to increase the science return of \HST\ through substantial increases in usable data,
either through higher signal-to-noise observations or more observations at the same signal to noise. As will be
shown, this objective has been accomplished.

It was well understood that a spectrograph with such a significant gain in sensitivity would enable numerous
other scientific endeavors, and that the community would identify science goals through the guest observer
(GO) process that we had not considered. Several science cases were identified by the COS science team,
including studying the origin of young stellar systems, stellar winds, cataclysmic variables, and active
galactic nuclei (AGN). However, it was the IGM science that drove the fundamental design choices for the FUV
optical design.  It is also important to recognize that COS was always assumed to be operating in parallel
with a functioning STIS. COS was not intended to replace, but rather to complement STIS by enabling
scientific observations that would otherwise require an unacceptably large commitment of observing time to
realize. For example, it was assumed that for high-resolution spectroscopy on bright targets, and high-precision imaging spectroscopy, STIS would remain the preferred instrument.

\begin{figure}
\epsscale{1.1}
 \plotone{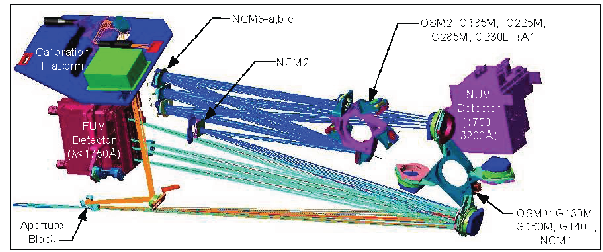}
 \figcaption{The COS optical layout showing the principal components of the instrument. Light from the telescope passeses through the aperture and strikes one of four optics on the Optics Slect Mechanism 1 (OSM1). In the FUV modes, light is corected and dispersed by one of three gratings (G130M, G150M or G140L) and directly imaged onto the FUV detector. If an NUV mode is selected, light is corrected and redirected by NUV Channel Mirror 1 (NCM1), collimated by NCM2, and then strikes one of five optics on Optics Select Mechanism 2 (OSM2).  The diffracted light is then re-imaged by the camera mirrors (NCM3-a,b,c) and recorded on the NUV MAMA detector.
 \label{CosOptPath}}
\end{figure}

The overall philosophy to maximize the sensitivity of COS centered on six fundamental design issues, which
are discussed below. For reference, a schematic representation of the COS FUV channel is shown in Figure~\ref{CosOptPath}.

\subsection{Minimize the total number of reflective surfaces in the optical design}\label{sec:minimize}

In the ultraviolet (UV), each reflective surface has at best 80\% reflectivity. Therefore the FUV channel was
designed to utilize a single optical element: one grating. This grating would have to disperse the light,
re-image the diverging telescope beam, correct the spherical aberration of the \emph{Hubble} Optical
Telescope Assembly (OTA) and correct the inherent aberrations of the quasi-Rowland circle mount.
The design of these gratings is discussed in Section~\ref{sec:ionetched}.

Previous \HST\ corrector designs for STIS \citep{STISdesign} and ACS \citep{ACSdesign} utilized multiple optics to correct the spherical aberration content of the OTA by using a collimating mirror to reform the entrance pupil (or equally, an image of the primary mirror) and then a corrector/re-imaging mirror to introduce
a wavefront correction of equal and opposite magnitude to the pathlength errors in the \HST\ primary mirror.
This is necessary when correcting the aberrations over a field. In such an application (e.g., STIS) a
slit can be introduced in the system at the corrected focus, and any post-slit spectrograph can utilize
classic designs which image the spectrograph entrance slit.  Such a design results in a large number of
reflective surfaces. However, by designing COS as a point-source spectrograph and limiting the field of
view to a few arc seconds, any point in the diverging beam behind the OTA focus is an effective pupil, and
the wavefront correction can be implemented at any point behind the focus. This is the technique utilized in COS.

\subsection {Eliminate the slit losses in the system}\label{sec:slitloss}

Most slitted spectrographs have significant throughput losses at the spectrograph entrance slit. Some
modes of STIS have up to 50\% losses at the slit \citep{Bost10}. The entrance slit in most spectrographs is present
to define the width of the imaged point source in the dispersion direction, to limit contamination from the
sky and other astronomical sources, and to relax requirements on pointing/tracking so that the spectral
resolution does not degrade if the telescope drifts on the sky. However, on \HST, these reasons are less
relevant. The excellent pointing stability of \HST\ (typically less than 0.007\arcsec\ jitter) eliminates
the need to correct for sky drift. The spherical aberration of the \HST\ OTA means that a point source
does not form a well defined image at any point without further correcting optics.
Therefore, any slit utilized
at the OTA focus in a classic spectrograph approach (where the final image is the image of the slit) will either
have low transmission if the slit is small, or poor spectral resolution if the slit is wide. COS avoids both of
these pitfalls by focusing the spectrograph not at the slit, but on the sky, so that spectral width of a
monochromatic point source is only as large as the corrected \HST\ line spread function (LSF) \citep{Kriss11} plus any aberrations
in the spectrograph. By proper design, these can be made small enough to meet the spectral resolution requirements.

\begin{figure}
 \plotone{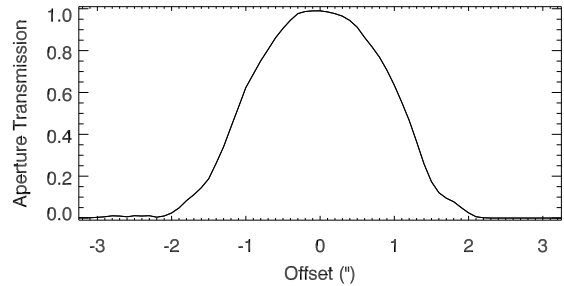}
 \figcaption{COS Primary Science Aperture (PSA) throughput versus offset in arc seconds along the dispersion direction.
 This on-orbit throughput measurement is derived from a large collection of target acquisition scans.\label{cosPSAthroughput}}
\end{figure}

As a consequence, the slit itself will be out of focus at the final spectral image plane, and the sky transmission curve
as a function of off-axis angle will not have a sharp cutoff (see Figure~\ref{cosPSAthroughput}). Effectively, COS is a slitless
spectrograph. However, to minimize sky contribution and potential confusion with nearby point sources, an
aperture is included at the minimum waist point in the \HST\ OTA point spread function. The minimum waist point
is not the same as the minimum encircled energy location, but rather the point at which 100\% of the light is
included in the minimum diameter. The diameter of the COS aperture is $\sim700 ~ \mu$m, which at the focal length of the
\HST\ OTA (57.6 m) corresponds to 2.5\arcsec\ in diameter. This slitless aspect of COS has implications
for the observation of diffuse sources, both in the \'{e}tendue ($A ~ \Omega$) of the instrument to diffuse emission, and to
the spectral resolution and spatial resolution of COS in such an application. These implications are discussed
in Section~\ref{sec:diffuse}.

Since the entrance aperture (effectively an out-of-focus field stop) is not at a location where the telescope
forms a real image (due to the spherical aberration), the COS spectrograph does not image the entrance slit,
but rather, forms an image of the sky at the spectral focal surface. (In a standard design where the entrance
slit is located in a stigmatic real image plane, this distinction is irrelevant.)  This is important to remember,
as objects outside the field of view of the aperture will pass light through the aperture due to the spherical
aberration, but the light will be reimaged by COS at the appropriate sky location, not the centroid of the
transmitted light in the aperture.

It should be noted that the \HST\ OTA, while excellent, is not perfect, and has aberration content due to figure
errors other than spherical aberration as well as scattered light from surface imperfections and dust. These
cause uncorrectable contributions to the \HST\ PSF which result in the COS slit passing slightly less than 100\%
of the original beam (approximately 98\%) and slightly degrading the final point spread function of the final
imaged light. These effects on the line spread function and the spectral resolution of COS are discussed in
Section~\ref{sec:spectralres}.

\subsection {Eliminate all transmitting optics in the system}\label{sec:transmitting}

Transmitting optics can introduce losses in the system, particularly at the shortest FUV wavelengths.
Additionally, the dominant source of dark noise in the systems utilizing windowed detectors is frequently
UV photons produced in the windows through phosphorescence.  This is true for the Multi-Anode Microchannel
Array (MAMA) detectors utilized by the COS NUV channels and the STIS FUV and NUV channels. By eliminating
windows in the detector systems, the throughput is enhanced and the noise level is greatly reduced, resulting
in large gains in signal to noise, particularly for the faintest targets with
$F_{\lambda} < 1.0 \times 10^{-14}$ \fluxunits\ at 1200 \AA.

Since the COS FUV channel has no transmitting optics at any point in the system, the front surface
reflectivity of the \HST\ OTA and FUV gratings provides measurable performance in the 900-1150~\AA~
region (and very limited performance at even shorter wavelengths), a unique capability on \HST.  This
performance is presented in detail in \citet{McCand10}. Becasue we could not be certain that the \HST\ OTA would provide any perfromance at these wavelengths,
the design of aberration control holography
did not consider these wavelengths, and the focal surface match of the detector is not maintained at the
extreme grating positions that must be employed to access these wavelengths. To have extended the design to these wavelengths would have necessitated reducing perfromance above 1150 \AA, and this seemed unwise given the possibility that the OTA might provide no reflectivity at these wavelengths.
Therefore, the spectral
resolution becomes substantially degraded at in these sub-1150~\AA\ modes, to a level of approximately R = 2000.
 Efforts are ongoing to optimize the resolution and efficiency in the sub-1150 \AA\ modes. However, since the \emph{Far
Ultraviolet Spectroscopic Explorer} (FUSE) \citep{Moos00} is no longer operational, COS provides
the only spectroscopic access to these wavelengths at the moment.

\subsection {Utilize ion-etched holographic gratings to minimize scattered light while maximizing net efficiency}\label{sec:ionetched}

\begin{figure}
	\plotone{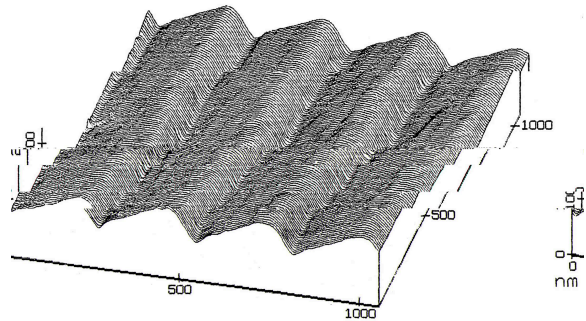}
	\figcaption{Measured FUV grating surface profile, achieved through ion etching of the hologrpahic master.  The line density is 3800 grooves mm$^{-1}$,
	and the blaze angle is approximately 14.3$^\circ$. \label{COSgrt}}
\end{figure}

The traditional trade-off between holographic gratings and mechanically ruled gratings is normally
considered to be one of high groove efficiency (mechanical) vs. low scatter (holographic). However,
proprietary techniques now exist that allow the introduction of a blazed groove profile on holographically ruled
gratings. Physical measurements of a COS FUV grating groove profile are shown in Figure~\ref{COSgrt}. This
allows the simultaneous development of a high groove efficiency and low scatter grating. To achieve such a profile first requires the exposure of the
holographic grating photoresist to a different depth than would be used for a sinusoidal grating, and then exposing the grating to an ion-ecth beam at multiple angles, each for a different exposure time. Not all combinations of line density and blaze angle can be reliably implemented.
The parameter space in optical design (line density, incoming angle, outgoing angle) for which the
appropriate blaze conditions can be physically implemented is constrained and not well defined
except through simulations of the ion-etch process. To simultaneously achieve the appropriate
spectral dispersion and have a manufacturable blaze angle, strict constraints are imposed on the
available optical design parameter space into which the spectrograph must conform, and even then it is necessary to iterate with the manufacturer to find a reliable solution.
Multiple test gratings were produced to optimze the flight production units, and it required 4-8 prototype runs to achieve optimal flight gratings.

The use of a holographic grating also permits the introduction of aberration control in the recording process,
allowing a single optic to maintain the performance over a significant bandpass. The actual design
of the gratings is presented in Table~\ref{FUVdesign}.

\begin{deluxetable*}{cccc}
\tabletypesize{\footnotesize}
\tablecaption{COS FUV spectrograph design parameters\label{FUVdesign}}
\tablewidth{0pt}
\tablehead{
  \colhead{Channel Name} &
  \colhead{G130M} &
  \colhead{G160M} &
  \colhead{G140L}
}
\startdata
HST Secondary Vertex/Slit (z) & 6414.4 mm & 6414.4 mm & 6414.4 mm \\
Slit Off Axis & 90.49 mm & 90.49 mm & 90.49 mm \\
Slit/Grating & 1652.57 mm & 1652.57 mm & 1652.57 mm \\
$\alpha$ & 20.1$^{\circ}$ & 20.1$^{\circ}$ & 7.40745$^{\circ}$ \\
$\beta$ & 8.6466$^{\circ}$ & 8.6466$^{\circ}$ & -4.04595$^{\circ}$ \\
$\alpha-\beta$ & 11.4534$^{\circ}$ & 11.4534$^{\circ}$ & 11.4534$^{\circ}$ \\
Grating/Detector & 1541.25 mm & 1541.25 mm & 1541.25 mm \\
Detector Normal/Central Ray & 9.04664$^{\circ}$ & 9.04664$^{\circ}$ & 9.04664$^{\circ}$ \\
Groove density (l/mm) & 3800. & 3093.3 & 480. \\
radius & 1652. mm & 1652. mm & 1613.87 mm \\
a4 & 1.45789E-9 & 1.45789E-9 & 1.33939E-9 \\
a6 & -4.85338E-15 & -4.85338E-15 & 1.48854E-13 \\
$\gamma$ & -71.0$^{\circ}$ & -62.5$^{\circ}$ & 10.0$^{\circ}$ \\
$\delta$ & 65.3512$^{\circ}$ & 38.5004$^{\circ}$ & 24.0722$^{\circ}$\\
rc & -4813.92 mm & -4363.6 mm & 3674.09 mm \\
rd & 5238.29 mm & 4180.27 mm & 3305.19 mm \\
recording $\lambda$ & 4880 \AA & 4880 \AA & 4880 \AA \\
\enddata
\end{deluxetable*}

The FUV G140L grating, which has much lower dispersion, was not ion-etched to create a triangular
groove profile, but utilizes a laminar profile instead. While this produces acceptable performance,
it is not as efficient as would be predicted with a triangular groove profile. Unfortunately, the
level of ion-etching required to produce a 1.4$^\circ$ blaze angle would remove so much material that
we could not guarantee the figure quality of the grating would be preserved. After the production of
the flight, laminar profile gratings, a test grating was produced with 1.4$^\circ$ blaze angles in late
2001.  However, it was never tested for figure quality, as the necessary \HST-simulating, spherical
aberration inducing optical feed system was needed at the end-to-end test level to support the scheduled
launch date of 2004.

\subsection {Utilize a large-format detector}\label{sec:largeformat}

Essential to the observational efficiency of a spectroscopic instrument like COS is the availability
of a large observational bandpass.  For IGM studies, the actual wavelength at which the absorbers
will appear is not known before the observation is made; therefore, complete coverage across the
entire 1150--1775~\AA\ bandpass in as few exposures as possible maximizes the net information
gathering capability of the instrument in fixed observing time. Traditionally, a large bandpass
is achieved through the use of an echelle spectrograph, such as on STIS. However, echelle spectrographs
require multiple gratings and optics, with the subsequent loss of throughput, and mechanically ruled echelle gratings
often exhibit significant scatter. (It has currently not been demonstrated that high-angle
triangular blaze profiles can be achieved through the ion-etching of holographic gratings.) Therefore,
it was decided to utilize a sufficiently large detector to obtain at least one half of the total
bandpass in a single exposure in first order. Fortunately, large microchannel plate detectors had
recently been developed for FUSE, and these served as the template
for the COS detectors. The active area for the COS detectors is two segments, each 85 mm long $\times$ 10 mm high,
curved to match the focal surface of the system, with a 9 mm gap between the segments. The readout system
for each segment is a cross delay line, resulting in the production of 16K $\times$ 1K non-physical pixels, which
do not arise from any physical boundaries, but from the digitization of a measured time interval. In this
way, the COS detectors are very different from a classical CCD or IR focal plane array, or even the STIS
MAMA detectors, whose pixel scale is established by a physical set of wires. For a much more complete
discussion of the detector development and capabilities please see \citet{COSgrdcalib}.

\subsection { Utilize an opaque photocathode}\label{sec:opaque}

Windowed detectors often utilize semi-transparent photocathodes applied to the back side of the windows.
However, higher detection quantum efficiencies (DQE) can normally be obtained by the use of opaque
photocathodes, which in the case of COS, are applied directly to the surface of the microchannel plates.
This has resulted in gain of up to four times the quantum efficiency for the photocathode compared to
semi-transparent applications. While the open-face design results in exposure of the photocathode to
the environment of the spectrograph, and subsequent degradation of the efficiency, current measurements
indicate a loss of approximately 5\% per year from the photocathode, which is small compared to the
significant gains realized through the opaque utilization.

\section{NUV Design Philosophy}\label{sec:NUVdesign}

The NUV channel was added to COS at the request of NASA immediately after its selection. It was intended
to provide a backup capability to the full STIS NUV wavelength regime (1750--3200 \AA) without significantly
increasing the overall cost or jeopardizing the performance of the FUV channels. Consistent with the low
cost approach, an existing flight qualified MAMA detector left over from the STIS program was dictated as
the detector to be utilized in the optical design. All of the photocathodes available for use in this regime
are required to be applied semi-transparently in a sealed tube for environmental protection, so that the FUV
detectors and optical concept were not applicable to the NUV design.
The STIS MAMA has a 25 mm active area (diameter) compared to the 170 mm active area of the COS FUV detectors.
STIS addressed this, utilizing an
echelle design to get the full NUV coverage in a single exposure. This approach also includes all the fundamental
limitations of an echelle design. It was decided that the COS design should not duplicate the STIS design, but
rather optimize the performance along a different figure of merit emphasizing maximum sensitivity over a limited
wavelength range. The figure of merit employed was for maximum sensitivity to a single absorption line that was narrower than the instrumental profile (15 km s$^{-1}$), and using the dispersion necessary to match the FWHM of the point spread function with three MAMA pixels. Further, we required that the design be capable of observing the entire wavelength range in multiple exposures. In this way, scientific objectives requiring only limited bandpass coverage (study of specific interstellar lines, for
example) could be implemented with higher sensitivity than in STIS, while science programs requiring the full
wavelength coverage would presumably be better executed with the STIS echelle modes. However, in the event of a STIS failure, the COS NUV channels would provide full wavelength coverage backup. A schematic representation
of the COS NUV channel is provided in Figure~\ref{CosOptPath}.

To keep the efficiency high and the scatter low, the NUV channel utilizes first-order holographic gratings.
In first order, at $R=20,000$ resolving power, the 25 mm of detector provides only limited wavelength
coverage. Therefore, the design required the introduction of a scanning mechanism to allow selection of any particular
wavelength region tuned to the observational requirements. Since the aberration control in a holographic design
is a function of the optical geometry, this meant that an aberration-free dispersive system had to be employed:
a flat grating with uniform, straight grooves operating in collimated light. This imposed the requirement that the beam be
corrected for spherical aberration and collimated prior to dispersion, and then re-imaged by a camera optic. Additionally, the longer
wavelength of the NUV channel necessitated a larger dispersion (in lines/mm) to accommodate the same resolving
power, $R~ = ~ 20,000$. In COS, the NUV channel mirror 1 (NCM1) corrects the spherical aberration, NCM2 re-collimates
the light, and the NCM3 mirrors re-image the light. In order to keep the optics a reasonable size, but maintain
the throw lengths necessary to accomplish the desired dispersion, the effective beam speed of the system is changed
from $f/24$ to $f/88$ by the NCM1/NCM2 combination. The increases the effective focal length by the same ratio,
creating an imaging plate scale at the spectral focal plane of 0.941\arcsec/mm or 0.023\arcsec\ per
25 $\mu$m pixel. It was decided to intercept the diffracted light with three camera systems (single optic re-imagers)
to maximize the total wavelength coverage in any single exposure. This results in three wavelength
segments that are non-contiguous in wavelength in any single exposure.
It is possible, with multiple exposures, to provide full, continuous wavelength coverage
over the full bandpass of any particular mode.

One of the locations on the NUV channel grating selection mechanism is occupied by a flat mirror, which results
in a broad-band NUV image forming at the MAMA surface. This channel is referred to a Target Acquisition 1
(TA1), and is primarily intended, as its name implies, for target acquisition purposes. However, it is available
for science applications (see Section~\ref{sec:TA1}).

\section{The Optical Design of COS}\label{sec:opticaldesign}

The FUV system includes two moderate resolution channels, G130M and G160M, centered on 130 and 160 nm respectively. There is also a low dispersion channel, G140L, centered at 140 nm. Because each the FUV channels incorporate a single optical element, the design of the FUV channel is simply the
design of the gratings. The gratings have a spherical substrate, combined with polynomial deformations in the
$4^{th}$ and $6^{th}$ orders to control the spherical aberration and approximate the inverse deviations of \HST\ primary
mirror from its designed hyperbolic shape. The holography is created with two point sources, one a virtual
source, and the nomenclature in the holographic recording geometry parameters follows the convention of Noda,
Namioka, \& Seya \citep{Noda74}. The elements listed in Table~\ref{FUVdesign} are defined below.

The \HST\ Secondary Vertex/Slit (z) is the distance from the vertex of the \HST\ OTA secondary to the COS primary science aperture, projected onto the optical axis of \HST\ (z).
Slit Off Axis is the position of the aperture perpendicular to the \HST\ optical axis. Slit/Grating is the distance from the aperture to the vertex of each grating. $\alpha$ and $\beta$ are the angles of the geometry as expressed in the standard grating equation: $\sin \alpha + \sin \beta ~ = ~ m \lambda / d$. The mount is in-plane and the ray along $\beta$ strikes the center of the detector. Grating/Detector is the distance from the grating vertex to the dtector focal surface along a ray leaving the grating center at an angle $\beta$ from the normal. Detector Normal/Central Ray is the tilt of the detctor relatrive to the primary ray along $\beta$. " Radius", a4 and a6 describe the surface of the grating as a sphere of radius "Radius" plus fourth-ordre ($ a4 ~ r^4 $) and sixth-order ($ a6 ~ r^6$ ) polynomial deviations from sphericity. $\gamma$, rc and $\delta$, rd are the  positions of the recording sources, measured in polar coordinates from the grating center and grating normal with the same sense as $\alpha$ and $\beta$. Note that rc is negative for G130M and G160M, indicating a virtual source on the opposite side of the grating as the optical surface. Recording $\lambda$ is the wavelength of the laser used for the holographic recording.  These parameters, along with the known surface description of the \HST\ OTA, and the radius of curvature of the detector surface, 826 mm, allows a full ray-trace simulation of COS FUV channles.

The optical layout of the COS FUV channels, at first glance, appears to be a Rowland circle. However, this
is not the case. The slit-to-grating distance in a Rowland circle is $R \cos (\alpha)$ and the grating-to-detector
distance is $R \cos (\beta)$ where $R$ is the diameter of the Rowland circle. The use of holography to correct the
astigmatism at more than one wavelength within the bandpass places tight constraints on the values of $\alpha$
and $\beta$ that can be employed in the design. Angles $\alpha$ and $\beta$ are further constrained in that $\alpha$-$\beta$
must fit within the physical dimensions available in the instrument, and the holographic recording sources must be
located in a physical configuration that will result in a high quality recording. All of these constraints combine
to require that $\alpha > \beta$. In a classic Rowland mount, this requirement is not constraining, and the slit-grating distance is less than the
grating-to-detector distance. An example of such a design is the \emph{Far Ultraviolet Spectroscopic Explorer} optical
design \citep{Green94}. However, in \HST, the slit location relative to the primary mirror is fixed, and
such a design would place the detector inside the \emph{Hubble} primary mirror,
and is clearly unacceptable. Therefore, as a starting point for the design, the two distances were swapped
(effectively, the object and image distances were interchanged, which under the thin lens approximation does
not alter the focal capability) and the resulting defocus and aberrations were corrected with holography.

The low-resolution FUV channel G140L presented a greater design challenge, as the beam angular deviation ($\alpha-\beta$) had
to be preserved, so as to make the dispersed light reach the detector, but the dispersion had to be 10\% of the
M mode dispersion. To achieve this, the grating had to be tilted substantially, creating large defocus terms
that had to be addressed by holographic correction. It was not possible to achieve a solution that corrected
the aberrations at two wavelengths in this geometry, which explains the larger astigmatism in the G140L channel. The design
parameters for the COS FUV spectrograph are given in Table~\ref{FUVdesign}.

All channels of the NUV channels utilize the same optical elements with the exception of the diffraction
grating (or mirror, in the case of TA1). NCM1 is the corrector and refocusing optic, NCM2 is the collimator
optic, and NCM3a, b and c are the three camera optics.

\section{Ground Performance}\label{sec:ground}

The stand-alone ground performance of the various components has been documented in the published literature.
For a discussion of the FUV gratings see \citet{Oste11a}. For details of the NUV components, refer
to \citet{Oste11b}. All components performed within specification except the NUV gratings.

The NUV gratings, as originally designed, provided substantially lower net effective throughput than anticipated.
These gratings are replicas of holographically ruled flats, with straight grooves, utilizing an optimized
pseudo-sinusoidal groove depth to maximize the efficiency into first order. When tested after replication,
with a platinum optical surface, they exhibited diffraction efficiencies as predicted. However, after final
optical coating with chromium (utilized as a diffusion barrier), aluminum and magnesium fluoride, all three medium-resolution (M) gratings performed well below specification. The G230L grating performed within specification.
Intense analysis was performed to understand the cause of the performance shortfall.  It was discovered that the
application of the magnesium fluoride overcoat altered the wavelength of the photons at the aluminum surface such
that the grating was operating at a diffraction anomaly \citep{Loew77} causing one polarization to have very low efficiency
into first order. It was decided to fly the G225M and G285M gratings with bare aluminum coatings. Bare aluminum
exhibits acceptable reflectivity at the wavelength in these bandpasses, and the literature indicated that after
an initial loss of reflectivity due to oxidation, the reflectivity would stabilize. The reflectivity of bare
aluminum in the bandpass of the G185M grating was considered unacceptable, and a MgF$_2$ coated G225M (with
subsequently lower line density) was substituted for the G185M grating. This moved the diffraction anomaly out of the
bandpass and provided an acceptable level of total efficiency. However, this also resulted in a lower dispersion
and spectral resolving power in the G185M channel, but at the time of this decision (2001) the imminent launch of
SM4, then scheduled for 2004, precluded the redesign and fabrication of new gratings.

Contrary to our expectations, the G225M and G285M gratings continued to exhibit losses in efficiency during the
storage of COS before launch. It had been anticipated that the efficiency loss would cease after 6 months to
1 year. Extensive testing of ruled witness samples ruled out contamination, metal diffusion, or groove profile
relaxation as sources of the continued efficiency loss. Bare aluminum witness samples (mirrors, not gratings) did
not exhibit reflectivity loss, even though they were stored with the flight gratings at all times, including inside the
instrument.  It was finally concluded that the most likely explanation was that the oxide layer
in the aluminum continued to grow on the surface of the gratings, even though they had beenstored under a nitrogen purge at
all times. This did not affect the reflectivity of the coating, but it did begin to create a dielectric coating on
the gratings that partially reintroduced the efficiency anomaly seen with the MgF$_2$ coating. Full solutions of
Maxwell's equations on the time-dependent modeled surface of the gratings, developed by the vendor, were able to reproduce the magnitude,
time-dependence and wavelength dependence of the efficiency losses in both gratings. Given the success of this
model, it was predicted that the efficiency losses would stabilize after the gratings were installed into \HST\ and
exposed to the vacuum of space. While some residual atomic oxygen does remain at \HST\ altitudes, at the very least,
it was anticipated that the degradation would slow down. However, the efficiency losses continue at the same rate
on orbit as observed on the ground during storage, 3.3\% per year for G225M and 10.8\% per year for G285M \citep{Ost11}.
At the time of this writing, we are at a loss to explain the data. However, given the limited utilization of these channels,
the additional exposure time needed to restore the science lost due to the efficiency loss is less than five orbits/year.

\section {Flight Performance}\label{sec:flight}

A description of the initial on-orbit performance of COS has been presented in \citet{Oste11a}. The flight
performance of the COS instrument is continually updated in the COS Instrument Handbook, available online at the
STScI website. The highlights of the performance are summarized here.

\subsection{FUV Throughput}\label{sec:FUVthroughput}

\begin{figure}[htb]
	\plotone{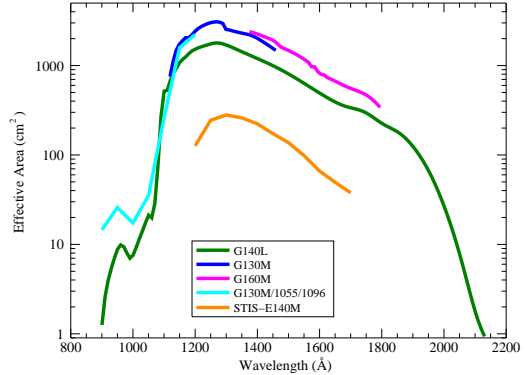}
	\figcaption{The measured, on-orbit effective area of the COS FUV channels as a function of wavelength.
	The nominal effective area for the STIS E140M channel is included for comparison. Slit transmission
	losses are not included for STIS.\label{cos_fuv_aeff}}
\end{figure}

The throughput on the FUV channels is excellent and exceeds the performance predictions in the proposal by
approximately a factor of two. This is the result of higher grating efficiencies than anticipated and higher
detector detection quantum efficiency (DQE) than assumed in the proposal. The on-orbit effective area of COS is
presented in Figure~\ref{cos_fuv_aeff}, along with the effective area of STIS in its most comparable mode.

\subsection{Backgrounds (dark and scattered light)}\label{sec:background}

\begin{figure}
	\plotone{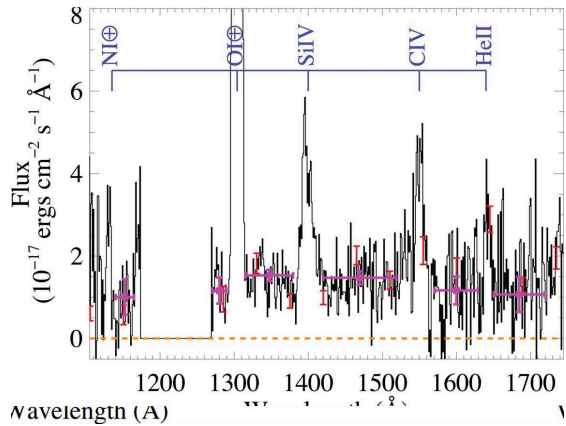}
	\figcaption{The time-averaged COS FUV spectrum of the black hole X-ray binary A0620-00 is shown in black
	from observations acquired on 23 March 2010. The spectrum has been binned to 2 resolution elements (15 pixels).
	The error bars shown in red are the statistical uncertainties from CALCOS propagated through the binning of the data points.
	Prominent emission features are labeled in blue (with airglow lines labeled with the circled plus signs).
	The purple bars show mean continuum fluxes, with the horizontal bars indicating the range over which the mean was
	calculated and the vertical bars showing the uncertainty of the mean.\label{fig_Froning10}}
\end{figure}

The on-orbit background for the FUV detector is approximately $\sim$1 count~s$^{-1}$~cm$^{-2}$, consistent with pre-launch
predictions \citep{Sah10}. The origin of this background is approximately $\sim$0.5 counts~s$^{-1}$~cm$^{-2}$ from internal radioactive decays in
the microchannel plate glass, and $\sim$0.5 counts~s$^{-1}$~cm$^{-2}$ from incident particles that the detector
electronics are unable to discriminate from real photon events. This results in $\sim0.8$ background
counts/resolution element in a nominal 2000 second ($\sim$1 orbit) observation. Being photon counting devices, the
COS detectors have zero read noise.  Except in the vicinity (within $\pm$2 \AA) of a bright airglow line, the
contribution from scattered light is even less than the dark background. This means that for brighter targets ($F_{\lambda} > 1.0 \times 10^{-13}$ \fluxunits)
the FUV channels are signal limited and have an effective dark background of zero. For fainter targets,
correction for background is necessary. However, the correction is much smaller than for previous \HST\
spectrographs. This allows COS to reach reasonable signal to noise at unprecedented low flux levels. One of
the goals stated in the proposal was to break through the $10^{-14}$ flux level ($10^{-14}$ \fluxunits)
which was considered the practical limit for observations with the existing and planned \HST\ instruments.
COS has measured line fluxes with equivalent continuum
fluxes of $1.0 \times 10^{-17}$ \fluxunits\ \citep{Fro11} such as the black hole X-ray binary seen
in Figure~\ref{fig_Froning10} and in the intergalactic \HeII\ absorption troughs (Figure~\ref{fig_Shull1}). Because of the extremely low background in
the FUV channel, even fainter fluxes could be observed if the science justified the exposure times necessary
to obtain the signal.

\begin{figure}
\plotone{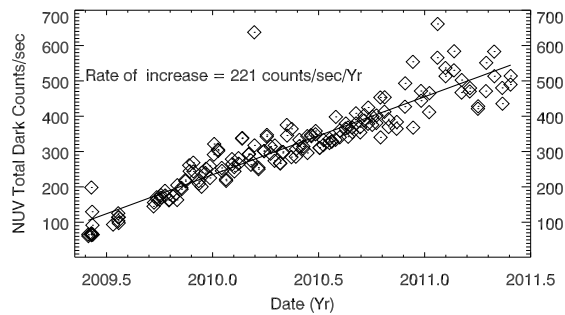}
\figcaption{On-orbit COS NUV detector dark rate shown for the entire detector.
A linear fit to the dark rate indicates a current rate of increase of 221 counts s$^{-1}$ yr$^{-1}$.\label{nuvdark}}
\end{figure}

The on-orbit background performance of the NUV MAMA is dominated by the phosphorescence of the detector window.
Ground testing of the COS detector, compared of the STIS flight detector, under similar conditions (utilizing
radioactive sources to stimulate the phosphorescence), indicated that the COS detector would have approximately
25\% of the dark rate of the STIS detector. As the STIS NUV detector had a dark rate of ~1200 counts s$^{-1}$ (over
the entire detector) before failure, it was anticipated that the COS MAMA would have $\sim300$ counts~s$^{-1}$
dark rate on-orbit. To our surprise, immediately after turn on, the count rate was only 60 counts s$^{-1}$. However, that
has been steadily rising on-orbit and is now close to 300 counts s$^{-1}$. The STIS NUV MAMA, on the other hand,
had a significantly enhanced background after repair, which has been steadily decreasing since re-commissioning.
The rates are displayed in Figure~\ref{nuvdark}. For a more detailed discussion, see \citet{Pen10}.

\subsection{Spectral Resolution}\label{sec:spectralres}
\begin{figure}
\plotone{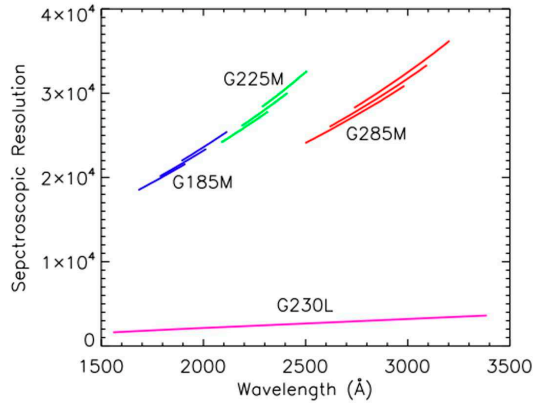}
\figcaption{NUV spectral resolving power ($R = {\lambda}/{\Delta \lambda}$) vs. wavelength as measured on-orbit. \label{nuvres}}
\end{figure}

The spectral resolving power of COS was originally designed to be $R \ge 20,000$ at all warformance is below this level. In the G185M (NUV) channel, the substitution of a lower
density grating has resulted in the resolving power ranging from 16,000--22,000 across the bandpass,
with the lowest resolving power at the shortest wavelengths. This reduction in spectral resolution was
accepted as preferable to the significantly reduced throughput that would have resulted from using the
original gratings. The resolving power of the NUV channels on orbit is presented in Figure~\ref{nuvres},
and is consistent with pre-launch expectations. The mid-frequency wavefront errors (MWFE) present in the \emph{Hubble}
primary and secondar mirrors which degrade the resolution in the FUV (see below)
modes is present in the NUV modes, but at a reduced level which does not reduce the resolving power below specification.

The FUV channels have lower resolving power on orbit than measured in thermal vacuum testing, due to
limitations of the \HST\ primary and secondary mirrors. Figure errors, primarily in the mid-frequency regime,
result in some power moving from the core to the wings of any \HST\ image. This effect is greater at shorter
wavelengths and is seen in all \HST\ instruments. It manifests as wings in optical images (at a significantly
reduced level due to the longer wavelengths involved), and in the spectrographs when sufficiently wide entrance
slits are employed. For example, this same effect is seen in STIS line profiles when the 2\arcsec\ slit is
employed.  The convolution of these aberrations with the COS FUV grating and FUV detector responses, results
in a line profile that is broader than a pure Gaussian with a FWHM equating to 1 part in 20,000
of the wavelength. Additionally, the depth of an absorption line is less from the true line spread function
than it would be if the COS LSF was a pure Gaussian.  Fortunately, the \HST\ OTA performance models \citep{Krist95}, when
convolved with the COS line spread function, as measured in ground testing, matches the on-orbit profile very
well \citep{Gha09}. In practice, this means that as long as profile fitting is done with the appropriate on-orbit
profiles, highly accurate parameters for the absorbing and/or emitting material can be derived. Except for
weak, narrow lines, the ability to constrain the physical parameters (as compared to the ideal line profile
capabilities) is limited by the signal to noise of the observation and not the degraded spectral resolving power.

As noted, this effect reduces the sensitivity of the instrument to detect weak ($< 10$ m\AA), narrow features
for a fixed signal to noise \citep{Keeney11}. It should be noted that for strong lines, or lines
significantly broader than the instrumental profile (greater than $\sim$50 \kms), this
phenomenon has virtually no effect on the ability to extract information from the spectrum at any signal-to-
noise level. Although the minimum detectable equivalent width, at a fixed signal to noise, is reduced from
this effect (compared to pre-launch predictions), the greatly increased signal to noise in COS spectra more
than makes up for this shortcoming, making COS vastly more sensitive to detecting weak features against a faint
continuum, in a fixed exposure time, when compared to STIS.

\begin{figure}
\plotone{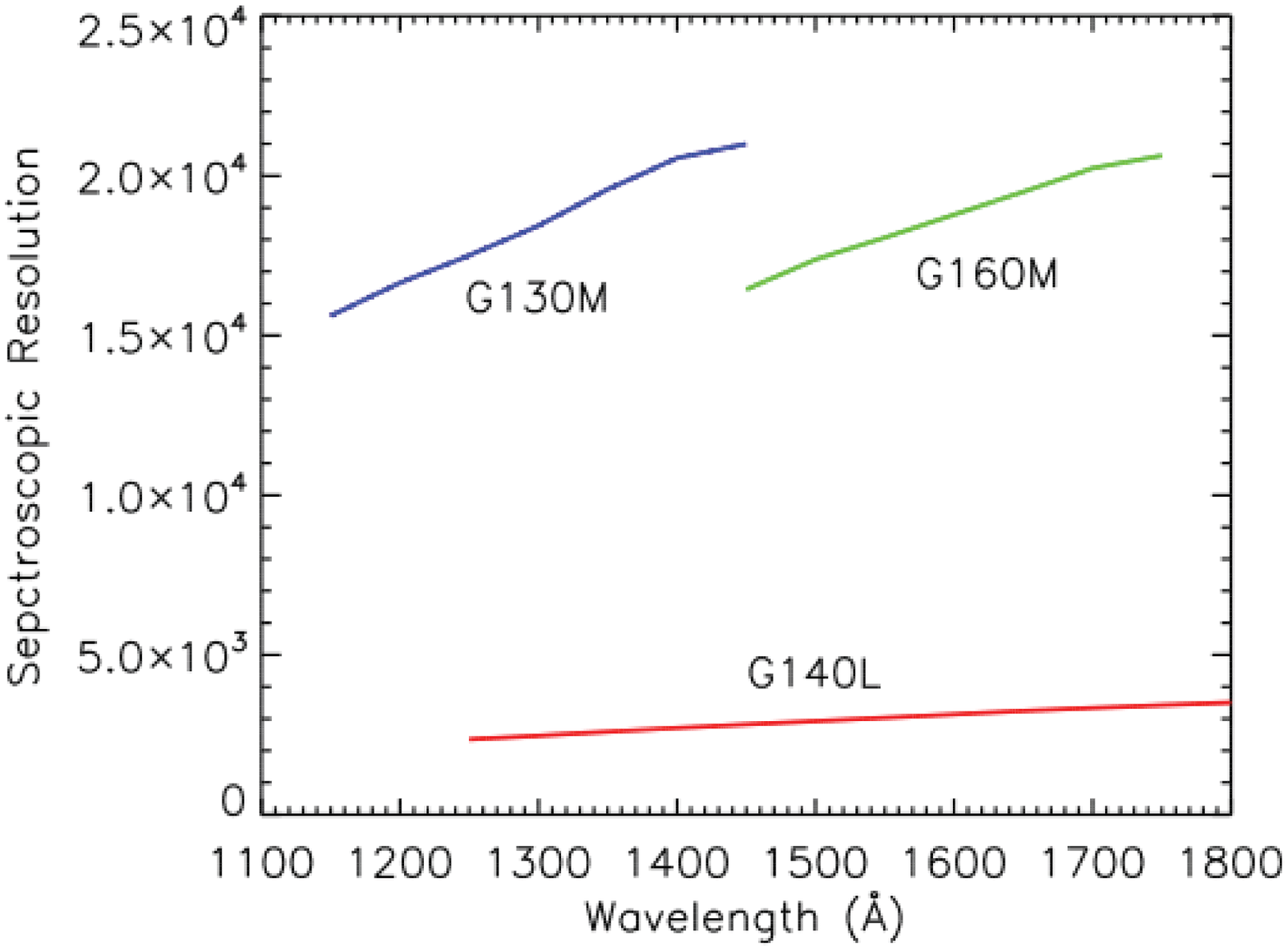}
\figcaption{FUV spectral resloving power ($R = {\lambda}/{\Delta \lambda}$) vs. wavelength as measured on orbit.\label{fuv_res}}
\end{figure}

Defining the spectral resolution ($\Delta\lambda$) of COS as the FWHM of the instrumental profile (in
physical coordinates) multiplied by the dispersion of the instrument yields a resolving power for COS
($\lambda/\Delta\lambda$) of 16,000 -- 22,000 in the FUV channels, with the highest resolving power at
the longest wavelengths in each channel. The resolving power increases primarily because $\lambda$ increases,
but also because the LSF improves at longer wavelengths (see Figure~\ref{fuv_res}).  For a detailed discussion of this
issue, refer to \citet{Kriss11}.

It was not originally anticipated that breakthrough science would result from COS observations due to an increase in resolving
power, since spectrographs with significantly higher resolving power have been (GHRS), and continue to be
available (STIS). However, the limited sensitivity of these instruments meant that many classes of astrophysical
targets could only be observed with their low-dispersion modes. Subsequent observations with COS have,
in fact, greatly increased the spectral resolution of the data on these targets, resulting in exciting new
science. As an example, in Figure~\ref{fig_TTauri} we show the spectrum of the classical T Tauri star AA Tau taken in 9900
s of COS observing time, and the previously best high signal-to-noise spectral data on this object, an ACS
prism spectrum with a 10,900 s exposure. Clearly the ability to understand the physical processes in the
target is greatly enhanced with the COS data, through its greatly improved spectral resolution and signal to noise.

\subsection{Spatial Resolution}\label{sec:spatialres}

Even though the design of COS FUV channel retains residual astigmatism in order to optimize the spectral
resolution with a single optic, spatial imaging at the approximately 1\arcsec\ level remains. Two point
sources, separated by 1\arcsec\ in the cross dispersion direction, will present completely separated spectra
in the FUV channel. At the time of this writing, this capability has not been utilized on any observations
that are beyond their proprietary period. However, recently executed observations of the inner ring of SN1987A (HST ID 12241, PI - R. Kirshner)
should be able to utilize this capability to great scientific benefit.

It should be noted that the calculation of the y-position (cross dispersion axis) by the detectors is done
independently of the x-position, and that this calculation should be corrected by the gain of each photon
event (the pulse height) to obtain the most accurate cross dispersion position and best imaging performance.
This y-position dependence on gain is referred to as y-walk and has only recently been well quantified
(Penton et al. 2012, in preparation). This should soon be implemented in the COS pipeline, as even for point sources, the
geometric correction of the detector involves both the x and y positions, and the flat-field corrections
are improved if this effect is properly accounted for.

In the NUV channel, the image is fully corrected by the use of multiple optics. Therefore, the imaging
capabilities are defined solely by the inherent aberrations of the correcting and camera mirrors, and the
detector resolution, which is the dominant effect. The cross dispersion resolution of NUV spectra is approximately
0.1\arcsec.

\subsection{TA1 imaging capabilities}\label{sec:TA1}

\begin{figure}
\plotone{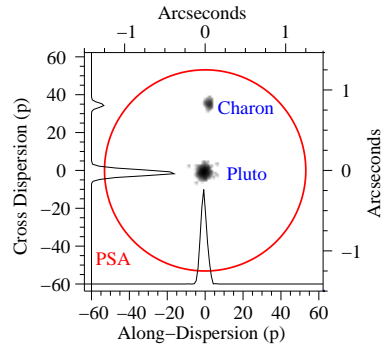}
\figcaption{TA1 image of Pluto and Charon taken for target acquisition. The angular separation of the objects is 0.8\arcsec.
The exposure time is 30 s. \label{TA1PlutoCharon}}
\end{figure}

When a flat mirror is placed into the optical path of the NUV channel instead of a diffraction grating,
the resulting mode is referred to as Target Acquisition 1 mode (TA1). This results in a highly sensitive
imaging mode with high angular resolution and a very limited field of view. In Figure~\ref{TA1PlutoCharon}
we show the image of Pluto and Charon, obtained in a 60 second exposure during target acquisition.
The separation of the two objects at the time of this exposure was 0.8\arcsec.

\subsection{Observing diffuse emission}\label{sec:diffuse}
\begin{figure}
\includegraphics[angle=90,scale=0.31]{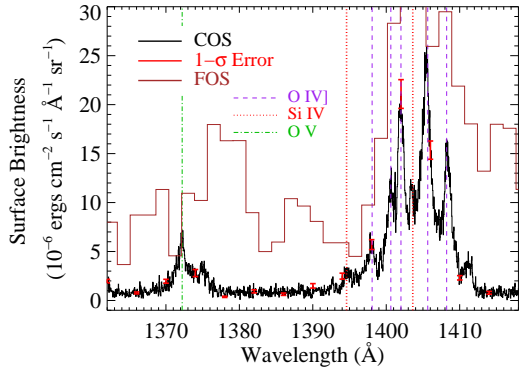}
\figcaption{COS spectra of the diffuse oxygen-rich supernova remnant N132D in the Large Magellanic Cloud adapted from~\citet{france09}.
The COS data are shown in black with representative error bars in red. Shock-excited oxygen and silicon ions are labeled. The previous deepest UV spectrum from
FOS is shown overplotted in brown. The COS data have been smoothed to three resolution elements for display, and the FOS spectrum has
been smoothed by three pixels. The spectral resolution is reduced for extended
source observations ($\Delta$$v$~$\approx$~200 km s$^{-1}$ with the FUV M-modes),
although COS provides large gains for low-resolution observations of faint, diffuse targets.}
 \label{fig_N132D}
\end{figure}

The 2.5\arcsec\ diameter aperture of COS makes the observation of diffuse objects highly sensitive
but with degraded spectral resolution.  Because the COS aperture rests at a point that is out of focus,
the actual sky acceptance of the aperture does not have a sharp edge, and light from regions outside the
nominal angular size of the COS aperture is accepted, while regions within the aperture, but near the edge,
are vignetted. Please refer to Figure~\ref{cosPSAthroughput}. Full illumination of the aperture reduces the resolving power of the M modes
to about 1500. The \'etendue ($A \Omega$) of COS at 1200 \AA\ is approximately $3 \times 10^{-7}$ cm$^{2}$~sr. As
an example of the ability of COS to observe diffuse emission, in Figure~\ref{fig_N132D} we show the early release
observations of the oxygen-rich LMC supernova remnant N132D (France et al. 2009), along with the previously best spectrum
obtained with the \HST\ Faint Object Spectrograph \citep{Blair00}.

\subsection{Sensitivity} \label{sec:sensitivity}

COS was designed to maximize the sensitivity of \HST\ with respect to UV spectroscopy. While comparisons
of instrument efficiency and effective area are informative, in the end it is observing time required to achieve a
particular signal to noise that must be considered the most direct comparison. A full signal-to-noise calculation
includes not only the collecting power of the instrument, but also the corrections to the final signal to
noise from dark backgrounds and scattered light.  For purposes of comparison to STIS, the times to achieve a
particular signal to noise in the COS G130M mode at 1200 \AA\ and at 1600 \AA\ are compared to that needed to
obtain a comparable signal to noise in the STIS E140M mode at the same wavelengths. The advantages of COS are
greater at the shortest wavelengths, as STIS scattered light performance and slit transmissions affect the final
signal to noise most powerfully at the shortest wavelengths, while the COS noise performance is relatively
wavelength independent. These comparisons also depend upon the nature of the spectrum observed. For example,
pure emission-line sources are less dependent upon correction for scattered light than observations of continuum
sources.  In the comparisons below, the source is assumed to be a flat continuum (in $F_{\lambda}$). Since the
STIS E140M mode has higher spectral resolution ($R = 42,000$) than the G130M mode, the comparison is also made at a S/N reduced by
$\sqrt{2}$ for STIS E140M, to reflect the ability to bin the higher resolution STIS data to have comparable
resolution to COS. It should also be noted that STIS E140M has a bandpass approximately twice the G130M bandpass,
so that if observation of the entire 1150--1750 \AA\ region is required, the COS net advantage needs to be
reduced by approximately a factor of 2. (All "time-to-signal-to-noise" calculations were made using the Space
Telescope Cycle 19 Exposure Time Calculator, with the STIS E140M
 $6\arcsec \times 0.2\arcsec$ slit, and default sky values.)

While COS has a 10--20 times advantage in effective area compared to STIS echelle modes, its greatly reduced
scattered-light levels and low backgrounds magnify its advantages at low flux levels, as can be seen in Table~\ref{ETCres}.
The original design goal of improving time to signal to noise by a factor of 10 has been substantially exceeded.
At the lower flux levels, S/N is dominated by corrections for noise rather than just by signal. The low scatter
and low background of COS are just as crucial for the data quality as the enhancements to the collecting power.

COS has broken through the $F_\lambda = 1\times10^{-14}$ \fluxunits\ flux barrier, reducing observing time for fainter systems by a factor of 100 or more, and making practical the systematic observation of hundreds of QSOís . Many Galactic targets of great interest also have flux levels at or below this
level. It is this enormous enhancement in observing power that has allowed COS to make revolutionary observations
of faint systems, both extragalactic and Galactic. Some of these observations are highlighted in the next section.

\begin{deluxetable*}{ccrrrrrrrrcrl}
\tabletypesize{\scriptnotesize}
\tablecaption{Results from Exposure Time Calculator for COS and STIS\label{ETCres}}
\tablehead{
\colhead{Flux (\fluxunits)} &
\colhead{Time to S/N: COS G130M/G160M} &
\colhead{Time to S/N: STIS E140M} &
\colhead{COS advantage} ]
}
\startdata
\multicolumn{4}{c}{\textbf{1200\AA}}\\
1.0E-13 & 1147 s for S/N = 30 & 66,879 s for S/N = 21.2 & 58 $\times$ faster \\
1.0E-14 & 5136 s for S/N = 20 & 578,049 s for S/N = 14.1 & 112 $\times$ faster \\
1.0E-15 & 13,766 s for S/N = 10 & 8.47e6 s for S/N = 7.1 & 615 $\times$ faster \\
\\\multicolumn{4}{c}{\textbf{1600\AA}}\\
1.0E-13 & 2279 s for S/N = 30 & 67,648 s for S/N = 21.2 & 30 $\times$ faster \\
1.0E-14 & 10,273 s for S/N = 20 & 776,717 s for S/N = 14.1 & 76  $\times$ faster \\
1.0E-15 & 29,330 s for S/N = 10 & 1.38e7 s for S/N = 7.1 & 470 $\times$ faster
\enddata
\end{deluxetable*}

\section{Science Highlights from COS Guaranteed Time Observations}\label{sec:ERO}
\subsection{COS High Signal-to-Noise IGM Observations}\label{sec:IGM}

At low redshift $\sim$90\% of the normal baryonic matter probably resides in the regions between galaxies
with only $\sim$10\% found in luminous objects (stars and galaxies) \citep{Fuku04}.  The cooler
30\% of the baryons with $T \approx (1-4)\times10^4$~K have been traced by the Ly$\alpha$ forest
\citep{Pen04,Leh07,Dan08} and some of the \ion{O}{6} lines \citep{Tripp08,Thom08}.
The warmer 20 to 30\% of the baryons with $T \approx (1-10) \times10^5$~K have
been traced by broad Ly$\alpha$ \citep{Rich06, Leh07,Dan10}, \ion{O}{6}
absorption \citep{Dan08,Tripp08} and Ne VIII absorption \citep{Sav05,Nara09}.
The remaining $\sim$40\% of the baryons probably reside in hotter gas with $T > 10^6$ K.
The hot baryons can be studied at X-ray wavelengths through searches for O VII and O VIII and other X-ray
absorption lines.  At UV wavelengths, warm and hot baryons can be studied via redshifted absorption lines
of Ne VIII, Mg X, Si XII, and \ion{O}{6} and by broad H I Ly$\alpha$ (BLA) absorption.

COS provides a major new facility for studies of the low redshift IGM. Although COS has a spectral
resolution lower than the E140M mode on the STIS, COS is $\sim$15 times more efficient
in collecting photons, making it possible to obtain relatively high S/N spectra of bright QSOs for IGM
absorption line studies. The search for Ne VIII and/or BLA absorption associated with \ion{O}{6} absorption is
greatly facilitated when the absorption line spectra have S/N $\sim$ 30 to 40.

The importance of the higher S/N spectra obtained by COS has been demonstrated through the detection
of Ne VIII ($\lambda\lambda770, 780$) absorption in an intervening multi-phase absorption system containing
O VI at $z$ = 0.495096 towards PKS~0405-123 ($z_{\rm em} = 0.5726$) by \citet{Nara10}. The detected
\ion{Ne}{8} directly points to high temperature gas since Ne VIII is very difficult to produce by photoionization
by the general extragalactic EUV radiation background.
Therefore, the \ion{Ne}{8} and associated \ion{O}{6} likely
arise in collisionally ionized gas at $T \approx 5 \times 10^5$~K with a large baryonic column density of
$\log N_H \approx 19- 20$. The multiphase absorber either traces the gas in the hot halo of a low luminosity
galaxy with an impact parameter of 110 $h_{70}^{-1}$~kpc detected by \citet{Chen09} or in a warm
IGM structure connecting to the galaxy.

\begin{figure}
\plotone{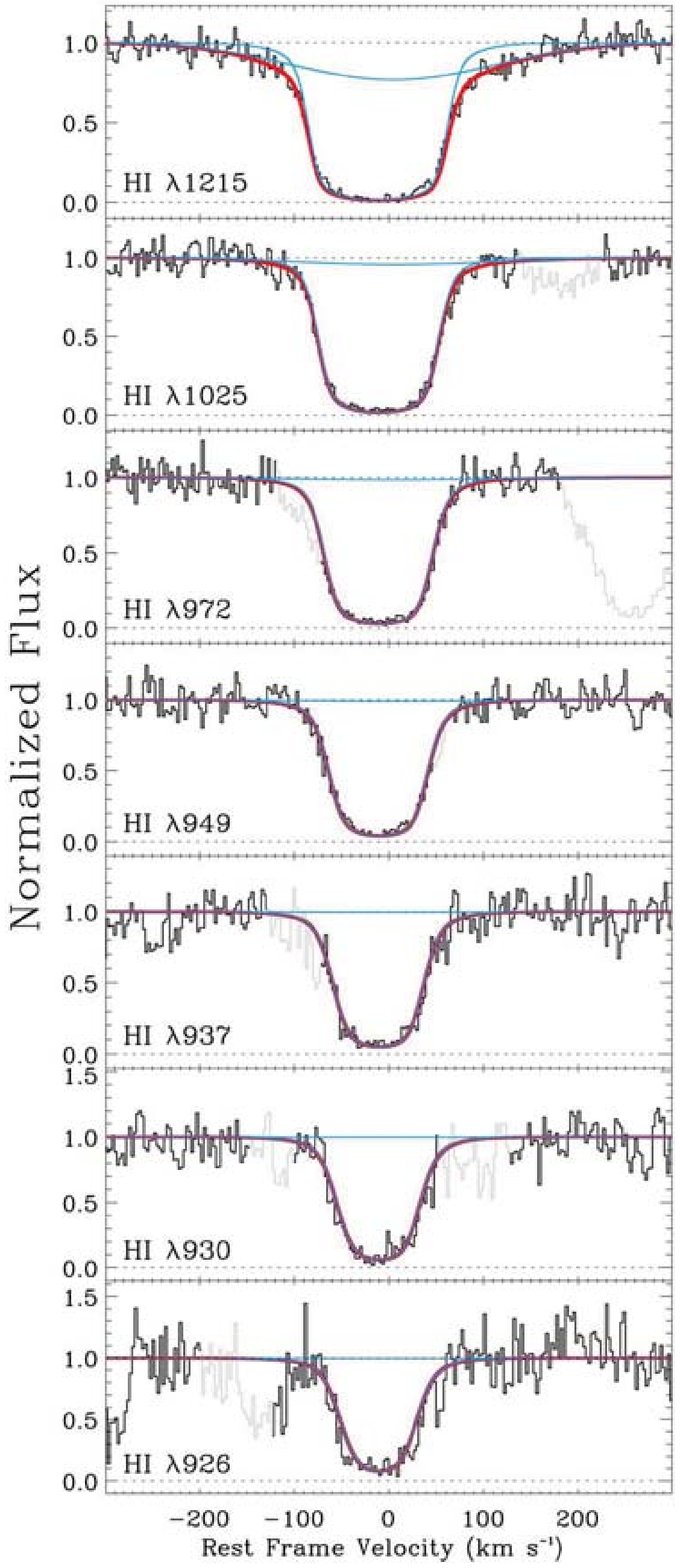}
\figcaption{Simultaneous two-component Voigt profile fits to the COS observations of {\HI} Lyman series (1216 to 926 \AA) absorption
at $z$ = 0.22601 in the spectrum of the QSO HE 0153-4110 from \citet{Sav11}. The {\HI} column density in the narrow {\HI} component was
determined to be log N({\HI}) = $16.61^{+0.12}_{ -0.17}$ from the Lyman Limit break in the FUSE observations of HE 0153-4520.
The broad wings on \Lya\ require the presence of a BLA with log N({\HI}) = $140^{+14}_{-16}$ {\kms} and log N ({\HI}) = $13.70^{+0.05}_{-0.08}$.
The BLA is associated with the presence of strong {\OVI} absorption in the multi-phase absorption system with log N({\OVI}) = $14.21\pm 0.02$
and b({\OVI}) = $37\pm1$ {\kms} and implies the direct detection of hot gas in the IGM with log T = $6.07^{+0.09}_{-0.12}$, [O/H] = $-0.28^{+0.09}_{-0.08}$
and log N(H) = $20.41^{+0.13}_{-0.17}$. \label{figBlair}}
\end{figure}

In another example, \citet{Sav11} have reported the detection of BLA and strong \ion{O}{6} absorption in
a multi-phase absorption system at $z$ = 0.22601 toward HE~0153-4520 ($z_{\rm em} = 0.450$). In this system,
a BLA with $b_{\rm HI} = 140^{+14}_{-16}$ km~s$^{-1}$ and $\log N_{\rm HI} = 13.70^{+0.05}_{-0.08}$ shown
in Figure~\ref{figBlair} is likely associated with the \ion{O}{6} having $b = 37 \pm 1$ km~s$^{-1}$ and $\log N_{\rm OVI} = 14.21 \pm 0.02$.
The \ion{O}{6} and BLA measurements imply the direct detection of a trace amount of thermally broadened H I in
hot gas with $\log T = 6.07^{+0.09}_{-0.12}$, [O/H] = $-0.28^{+0.09}_{-0.08}$, and $\log N_{\rm H} = 20.41^{+0.13}_{-0.17}$.
The observation illustrates the power of COS to detect and study gas in the low redshift IGM with $\log T \approx 6$
and a very large total column density of hydrogen.  The observed reference to H I via the BLA is crucial for
determining the temperature, metallicity and total baryonic content of the absorber.

Additional examples of the power of high S/N IGM observations with COS include the detection of an
H I-free broad \ion{O}{6} absorber at $z$ = 0.1670 toward PKS~0405-124 \citep{Sav10} revealing hot
gas associated with a pair of galaxies and the BLA and \ion{O}{6} absorber tracing a galaxy filament at
$z = 0.01025$ toward Mrk 290, where $\log T \approx 5.14$, $\log N_{\rm H} \approx 19.6$, and
[O/H] $\approx -1.7$ \citep{Nara10}.

\subsection{The \ion{He}{2} Reionization Epoch}\label{sec:HeII}
Determining when and how the universe was ionized has been an important question
in cosmology for decades \citep{GP65,Fan06,Meik09}.
The epoch of reionization in hydrogen is a probe of the transition from neutral to ionized hydrogen
in the intergalactic medium (IGM) that occurred between redshifts $z =$ 7--12, marking
the exit from the cosmic ``dark ages". Helium underwent similar reionization from \ion{He}{2} to \ion{He}{3}
at $z = 2.8 \pm 0.2$ \citep{Reim97,Kriss01,Shull04}, likely mediated by
the harder ($E \geq 54.4$ eV) radiation from quasars and other active galactic nuclei (AGN).
With a 4 ryd ionization potential, He$^+$ is harder to ionize than H$^0$, and He$^{+2}$ and recombines
5--6 times faster than H$^+$ \citep{far98}.  For these reasons, and the fact that
most hot stars lack strong 4 ryd continua, AGN are likely to be the primary agents of \ion{He}{2}
reionization.  Owing to its resilience, He$^+$ is much more abundant than H$^0$, with predicted
column-density ratios $\eta \equiv$ N(\HeII)/N(\HI) $\approx$ 50--100
\citep{Mir96,far98}.

\begin{figure}[htb]
\plotone{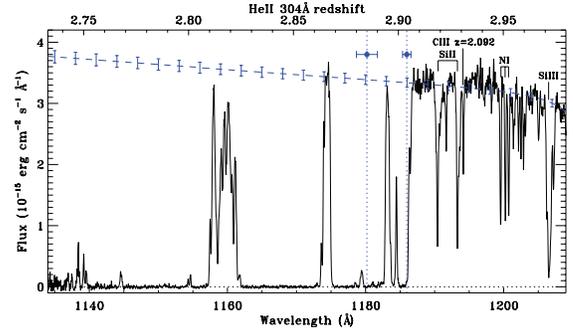}
\figcaption{COS/G130M data (Shull \etal\ 2010) showing \HeII\ absorption troughs and
  flux transmission windows at redshifts $z \leq 2.9$. The proposed QSO systemic redshifts
  ($z_{\rm QSO} = 2.885$ and 2.904) and extrapolated AGN continuum are marked. Interstellar lines
  of \NI, \SiII, \SiIII\ appear longward of 1190~\AA, and strong \HeII\ absorption is seen shortward of
  1186.26 \AA, with windows of flux transmission near 1183, 1174, 1160, and 1139~\AA. Inexplicably,
  no QSO-proximity effect is seen in flux transmission near $z =$ 2.86--2.90.
  Transmission recovers at $\lambda < 1100$~\AA, shortward of three troughs of strong \HeII\ absorption
  ($\tau_{\rm HeII} \geq 5$) that span very large redshift intervals (and comoving radial distances) between
  $z =$ 2.751--2.807 (61 Mpc), $z =$ 2.823--2.860 (39 Mpc), and $z =$ 2.868--2.892 (25 Mpc). \label{fig_Shull1}}
\end{figure}

HST/COS obtained UV spectra \citep{Shull10} of the bright ($V = 16$) high-redshift
($z_{\rm em} \approx 2.9$) quasar, HE~2347$-$4342, using both the G130M (medium-resolution,
1135--1440~\AA) and G140L (low-resolution, 1030--2000~\AA) gratings. The spectra exhibit patchy
absorption in the \Lya\ line of \ion{He}{2} $\lambda303.78$ (Figure \ref{fig_Shull1}) between $z =$ 2.39--2.87 (G140L) and
$z =$ 2.74--2.90 (G130M).  COS provides better spectral resolution, higher-S/N, and much better determined
backgrounds than previous studies, with sensitivity to abundance fractions $x_{\rm HeII} \approx 0.01$
in low-density filaments of the cosmic web. The COS spectra provide clear evidence that the epoch of
\ion{He}{2} reionization is delayed below $z = 3$, with intervals of patchy \ion{He}{2} flux-transmission, punctuated by
three long (5--10 \AA) troughs of strong \ion{He}{2} absorption. These troughs, spanning 25--60 Mpc of proper
distance between $z = $2.751--2.807, $z =$ 2.823--2.860, and $z =$ 2.868--2.892, probably arise from
filaments in the baryon distribution and the scarcity of any strong photoionizing source (AGN) within 30--50 Mpc.
They are uncharacteristic \citep{Dix09} of the IGM if \ion{He}{2} was reionized at $z \approx 3$,
and they suggest that the epoch of \ion{He}{2} reionization extends to $z \la 2.7$.  A comparison of \ion{He}{2} and \HI\
absorption in \Lya\ shows optical-depth fluctuations in the \HeII/\HI\ ratio on scales
$\Delta z \approx$ 0.003--0.01.  Corresponding to 4--10 Mpc in comoving distance, these fluctuations likely
arise from AGN source variations (1--4 ryd spectral indices) and IGM radiative transfer effects. Probes of the
\ion{He}{2} abundance fraction, $x_{\rm HeII} \geq (10^{-2})(0.1/\delta_{\rm He})(\tau_{\rm HeII} /5)$, at redshifts $z = 2.7-3.0$
are now possible at optical depths $\tau_{\rm HeII} \geq 5$ in low-density regions (over-densities
$\delta_{\rm He} \approx 0.1$). The \ion{He}{2} absorption is 50--100 times more sensitive to trace abundances than
\HI\ at $z = 6$, where a neutral fraction $x_{\rm HI} = 10^{-4}$ saturates the \Lya\ absorption.

\subsection{Galactic High-Velocity Clouds}\label{sec:HVC}

Absorption spectra in the UV provide sensitive diagnostics of conditions in the halo of the Milky Way,
where competing process of accretion and outflow determine the evolution of the Galaxy.  Recent \HST\ spectra
provide considerable insight into the infall of low-metallicity gas onto the disk, an ongoing process that can account
for the observed stellar metallicities, star formation rates, and mass-metallicity relations \citep{Pag94}.
A previous \HST/STIS survey of Galactic high-velocity clouds (HVCs) in the sensitive UV
absorption line of \ion{Si}{3} $\lambda1206.50$ \AA\ \citep{Shull09} suggested that the low Galactic halo is enveloped by
a sheath of ionized, low-metallicity gas. This gas reservoir can provide a substantial cooling inflow
($\sim1~M_{\odot}$ yr$^{-1}$) to help replenish star formation in the Galactic disk, recently estimated at
2--4 $M_{\odot}$ yr$^{-1}$ \citep{Diehl96,Rob10}.  Galactic disk formation is believed to
occur by the gradual accretion of pristine or partially processed material into the interstellar medium.  The
metallicity of the initial reservoir of gas is enriched by ejecta from star formation, mixed with infalling low-metallicity
gas, possibly through ``cold-mode accretion" \citep{Dek06,Ker09}. This process continues
to the present day, regulated in a manner that produces the local G-dwarf metallicity distribution and avoids the
overproduction of metal-poor disk stars. The infall of gas from the intergalactic medium and Galactic halo
also places a chemical imprint on mass--metallicity relations \citep{Erb06}.

\begin{figure}
\plotone{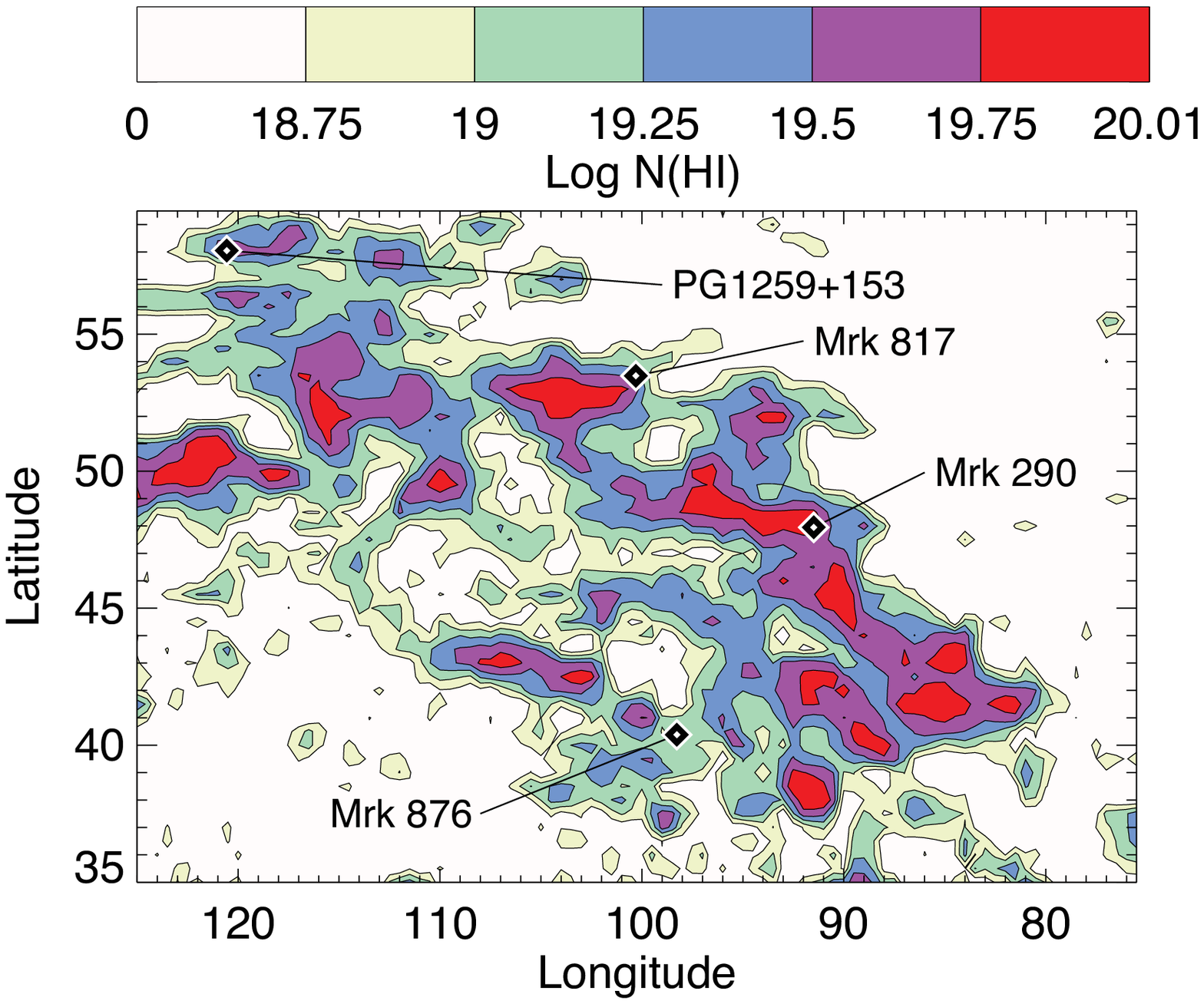}
\figcaption {Locations of four AGN targets studied by COS \citep{Shull11} overlaid on
  contour map of 21-cm emission from the Leiden-Argentine-Bonn (LAB) survey \citep{Kalb05}
  with $\sim0.6^{\circ}$ angular resolution on a $0.5^{\circ}$ grid in $\ell$ and $b$.
  Emission is shown over Galactic coordinates $\ell = 75-125^{\circ}$
  and $b = 35-60^{\circ}$ and Complex-C velocities between $-210$ and $-95$ \kms.
  At the 10 kpc distance of Complex~C, 1 degree corresponds to 175 pc. \label{fig_Shull2}}
\end{figure}

COS recently obtained high-quality UV spectra of HVCs in Complex~M \citep{Yao11} and Complex C
\citep{Shull11}. The wide spectral coverage and higher S/N, compared to pre-SM4 \HST\ spectra, provide
better velocity definition of the HVC absorption, additional ionization species, and improved abundances in halo gas.
Complex~M abundances are high, between 0.85--3.5~$Z_{\odot}$, suggesting that it consists of returning
Galactic fountain gas. In contrast, Complex C has a low metallicity (0.1--0.3 $Z_{\odot}$; \citet{CSG07})
and a wide range of ion states that suggest dynamical and thermal interactions with hot gas in the Galactic halo. COS
studied 4 AGN sight lines (Figure~\ref{fig_Shull2}) toward Mrk~817, Mrk~290, Mrk~876, and PG~1259+593. Spectra in the COS
 medium-resolution G130M (1133--1468~\AA) and G160M (1383--1796~\AA) gratings detect UV absorption
 lines from 8 elements in low ionization stages (\OI, \NI, \CII, \SII, \SiII, \AlII, \FeII, \PII) and 3 elements in intermediate
 and high-ionization states (\SiIII, \SiIV, \CIV, \NV).  The HVCs exhibit a range of \HI\ and \OVI\ column densities,
$\log N_{\rm HI} =$ 19.39--20.05 and $\log N_{\rm OVI} =$ 13.58--14.10, with substantial amounts of kinematically
associated photoionized gas. The high-ion abundance ratios are consistent with models of cooling interfaces between
photoionized and collisionally ionized gas: N(\CIV)/N(\OVI) $\approx$ 0.3--0.5,
N(\SiIV)/N(\OVI) $\approx$ 0.05--0.11, N(\NV)/N(\OVI) $\approx$ 0.07--0.13, and N(\SiIV)/N(\SiIII) $\approx 0.2$.
The thermal pressure of the hot Galactic halo, $P_{\rm halo}/k \approx $ 10-20~cm$^{-3}~{\rm K}$, is
comparable to the ram pressure on the HVCs. As the HVCs encounter higher densities in the stratified lower halo,
they may be dissipated by interface instabilities, feeding the Galactic halo rather than replenishing
the reservoir of star formation in the disk.  Individual clumps of Complex~C (Figure~\ref{fig_Shull2}) appear to be near virial
equilibrium with pressure confinement. Their masses ($\geq10^5\,M_{\odot}$) are probably above the threshold
for survival against interface instabilities (Heitsch \& Putman 2009).

\subsection{New Molecular Tracers for Protoplanetary Disk Studies}\label{sec:disk}

The combination of high-sensitivity, low instrumental backgrounds, and moderate spectral resolution afforded by
COS has provided a new window onto the physical and chemical state of protoplanetary disks. Previously, only a
small sample of the nearest young (age $\lesssim10$~Myr) low-mass circumstellar disk targets were bright enough
for moderate resolution FUV spectroscopy \citep{Herc06}. The most extensive surveys
UV were carried out with the G140L mode of STIS and the spectroscopic modes on ACS (e.g. the PR130L
mode, \citet{Ing09}). The large instrumental background and very low dispersion of these data severely
limits their utility for measurments of line identifications, fluxes, and widths. We are in the process of
assembling a large sample ($>$ 30) of protoplanetary disks systems by combining the GTO targets with several
GO programs. This work is shedding new light on the molecular composition and physical characteristics of the
planet-forming regions ($a \lesssim 5$~AU) in these systems.

Photo-excited molecular hydrogen (H$_2$) emission has long been recognized as an important probe of
protoplanetary disk surfaces \citep{Brown81,Val00}, however low-resolution and/or high
instrumental backgrounds have prevented a detailed analysis of this emission in all but the brightest
sources \citep{Herc02,Herc04,Walter03}. With COS, we are not only acquiring high signal-to-noise,
velocity-resolved profiles of tens to hundreds of H$_{2}$ emission lines in $every$ gas-rich disk observed
(Yang et al. 2011), we have begun to push these studies into the regime of protoplanetary disks orbiting
brown dwarfs \citep{France10a}. The ability to separate line and continuum features at very faint flux
levels ($F_{\lambda}$~$\lesssim$~2~$\times$~10$^{-15}$ \fluxunits) has enabled the
first detailed study of H$_{2}$ excited by non-thermal electrons \citep{France11a} in a protoplanetary
environment.

\begin{figure}
 \plotone{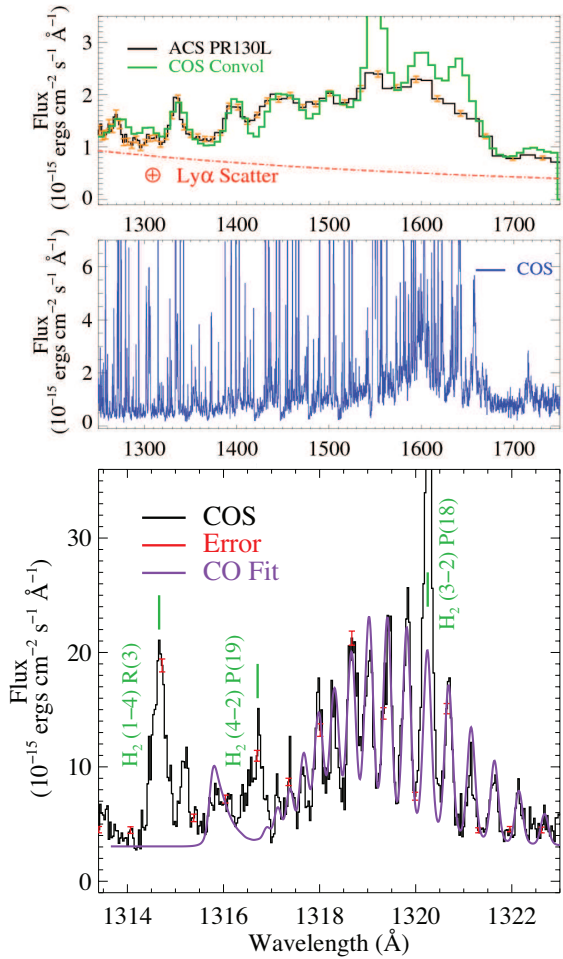}
 \figcaption {Top: The ACS PR130L spectrum of the T Tauri star AA Tau \citep{Ing09} (exposure time: 10,900 s)
 and the COS spectrum convolved through the ACS prism line spread function. Middle: The COS spectrum of AA Tau
 (exposure time: 9900 s) over the same bandpass. Bottom: An expanded view of the COS spectrum of another T Tauri star,
 DF Tau, including a model fit for CO emission. \label{fig_TTauri}}
\end{figure}

Perhaps the most intriguing discovery in the COS circumstellar disk program is the detection and
characterization of carbon monoxide (CO) in FUV spectra of low-mass protoplanetary systems.
\citet{France11b} presented the first spectrally resolved detections of emission and absorption of
lines of CO in these environments. This work showed that CO can account for a significant fraction of
the ``1600~\AA\ bump'' previously thought to be H$_{2}$ emission that has been suggested as a potential
measure of the mass surface density of protoplanetary disks. Initial modeling results (Figure~\ref{fig_TTauri};
\citet{Schin11}) indicate that the photo-excited CO population in these disks may be spatially
distinct from the well-studied $M$-band CO fundamental emission (see e.g., \citep{Naji07}), suggesting
a stratification of the molecular structure in low-mass protoplanetary disks not predicted by existing models.

\subsection{Interstellar Observations of Translucent Clouds} \label{sec:translucent}

The high sensitivity and low background of COS has allowed for the
observation through heavily reddened sightlines. Five stars
have been observed, with A$_V$ ranging from 2.7--7.6. Such a
high-extinction sightline may be simply a long sightline through
low-density clouds, as in Cyg OB2 8A (A$_V$ = 4.83; \citet{Snow10}),
which shows atomic depletions and molecular content consistent with
the diffuse interstellar medium, or may sample translucent clouds,
which represent a change in photochemical regimes due to higher
densities and heavy shielding from the photodissociating interstellar
radiation field \citep{Shef08}. In the latter case, carbon
monoxide (CO) becomes the dominant form of gas-phase carbon, and
\citet{Burgh07} showed that the CO/H$_2$ ratio provides a good indicator
of the presence of translucent material.

One sightline in particular may have even probed dense cloud material,
an unprecedented look at a regime formerly only accessible to infrared
or longer wavelengths. NGC~2024-1 ($A_V = 7.6$) exhibits a CH column
density indicating a molecular fraction near unity and some of the
strongest UV CO absorptions ever observed (Snow et al., in
preparation). The inferred CO/H$_2$ ratio is near 10$^{-4}$, indicating that
nearly all the carbon along the line of sight is in CO. In
conjunction with UV absorption spectroscopy taken previously,
COS allows us to probe the transition region between the diffuse and
the dark clouds in a consistent manner.

\subsection{Star Forming Regions}\label{sec:SFR}

The star-formation rate densities in starburst galaxies can exceed the values found in normal
galaxies by orders of magnitude. The ionizing radiation from the newly formed young stars can
lead to prominent Ly$\alpha$ emission due to recombination of hydrogen in the ambient interstellar
medium. The Ly$\alpha$ line has become be an important spectral signature in young galaxies at
high redshift since the Ly$\alpha$ luminosity amounts to a few percent of the total galaxy
luminosity concentrated in a small wavelength bin \citep{Stia04}. However,
the assumption of Ly$\alpha$ being a pure recombination line in a gaseous medium is too simple.
Resonant scattering by atomic hydrogen significantly increases the likelihood of Ly$\alpha$
destruction by dust, leading to much lower line strengths \citep{Neu90,Char93}.
Equally important, the kinematic properties of the interstellar medium may very well be the
dominant escape or trapping mechanism for Ly$\alpha$ \citep{Mas03}.

The theoretical situation is sufficiently complex that observational tests are called for.
While a large body of data exists at high-$z$ (\citet{Hayes10}, and references therein),
there are few (if any) Ly$\alpha$ surveys of starburst galaxies at low-$z$. Such
surveys are motivated by the claim of a trend of increasing Ly$\alpha$ escape fraction with redshift
\citep{Deha08}. Moreover, spectroscopy of local Ly$\alpha$ emitters can be
complemented by high-resolution imagery, thus providing additional constraints not available
at high redshift. Previous generations of UV spectrographs were quickly pushed beyond their
limits when observing suitable candidate galaxies; in order to be able to observe Ly$\alpha$
intrinsic to the galaxy, the redshifts should be high enough to avoid Galactic foreground
Ly$\alpha$ absorption, which at the same time implies large distances and faint flux
levels in the UV.

The vastly superior sensitivity has finally enabled observations of local starburst galaxies
at large enough redshift and in statistically meaningful numbers to study their Ly$\alpha$
properties. We selected a sample of 20 star-forming galaxies from the KISS database
\citep{Salzer00}. These galaxies form a homogeneous set selected via their H$\alpha$
emission, with a wide range of oxygen abundances, dust reddening, and luminosities.

\begin{figure*}[htb]
\begin{center}
\includegraphics[angle=90,scale=0.6]{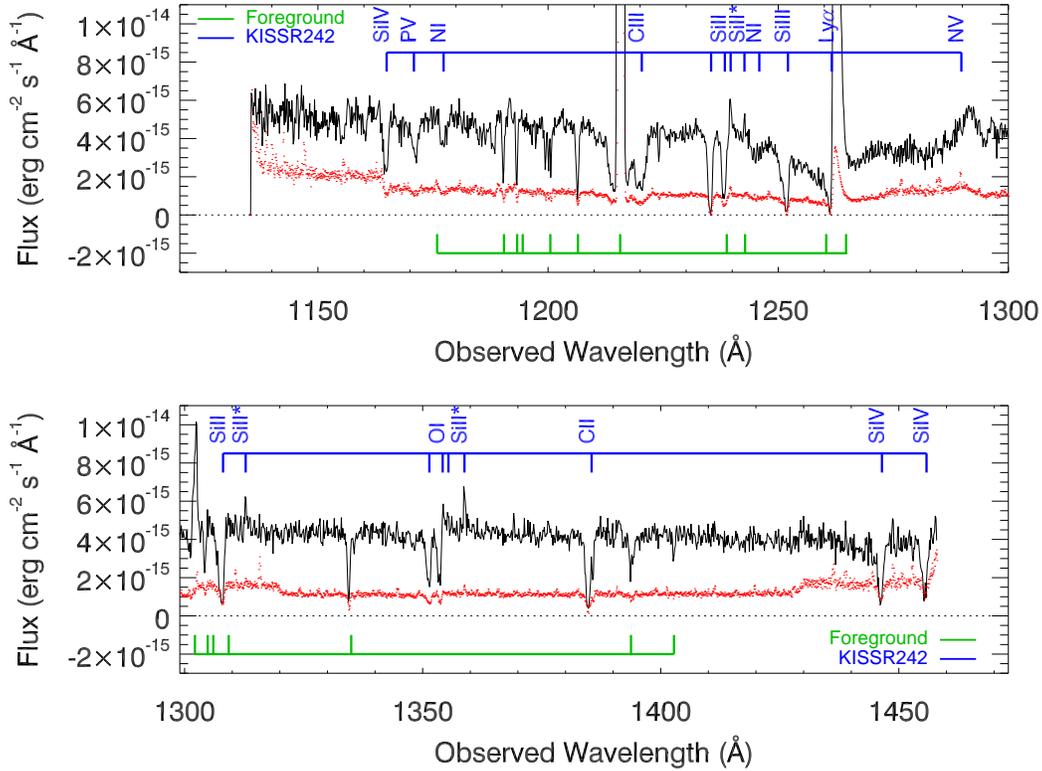}
\end{center}
\figcaption{COS G130M rest-frame spectrum of the H$\alpha$ selected galaxy KISSR0242, taken from the KISS database \citep{Salzer00}.
This spectrum (adapted from France et al. 2010b) was obtained in a single orbit using the G130M grating at two central wavelengths.
The data are shown in black, with the error vector overplotted as the dotted red line.
Foreground absorbers are identified with green tickmarks and the emission and absorption features
at the redshift of KISSR0242 are labeled in blue.\label{fig_StarForm}}
\end{figure*}

In Figure~\ref{fig_StarForm} we show COS G130M spectra obtained for two of the program
galaxies. The
plethora of spectral features can be grouped into broad stellar-wind lines
(e.g., N~.V $\lambda$1240), numerous weak stellar photospheric absorptions, narrow
interstellar lines (e.g., C~II $\lambda$1335), and the complex Ly$\alpha$ feature
\citep{Lei11}. Ly$\alpha$ can be a pure, damped profile, an emission line
associated with a galactic outflow, or a combination of both. The rich detail provided
by COS will allow us to study the physical conditions in the Ly$\alpha$ emitting
regions (including the Ly$\alpha$ escape fraction)and their correlations with, e.g.,
the galaxy kinematics, dust content, and star-formation properties. Early results from these efforts are presented in \citet{france10b}.

\section{Conclusions}\label{sec:conclusions}

The Cosmic Origins Spectrograph has been successfully installed in the \emph{Hubble Space
Telescope} and is returning high quality scientific data on a wide variety of astrophysical
targets. It has achieved its primary objective of vastly increasing the sensitivity of
UV spectroscopy on \HST, by a factor of 100 or more for faint targets, and opened
up new observational opportunities for the astronomical community.

\acknowledgments

The authors would like to acknowledge the enitire team at Goddard Sapce Flight Center, Ball Aerospace, JY Horiba, and the Space Telescope Science Institute for making the Cosmic Origins Spectrograph a reality. Literally hundreds of people contributed to its success. We especially thank Hsiao Smith, Jean Flammand, Francis Bonnemason and Francis Cepollina for their years of service to the program. We also thank Gregory Herczeg for providing the AA Tau data. This work was supported by NASA programs NAS5-98043, NAG5-12279 and NNX08AC14G.


\end{document}